\documentclass[letterpaper, 11pt]{article}

\usepackage[pdfusetitle]{hyperref}
\usepackage[margin=1in, top = 0.5in, bottom = 0.5in]{geometry}
\usepackage{framed}

\begin{document}

\title{Split block Bloom filters}
\author{Jim Apple \\
  jbapple@jbapple.com}
\maketitle
\thispagestyle{empty}

Bloom filters and other approximate membership query structures (including quotient filters and cuckoo filters) typically have worst-case operation costs that are linear (or worse) in $\lg (1/\varepsilon)$. \cite{cuckoo-filter,quotient-filter}
In Bloom filters, this is driven by the $\lg (1/\varepsilon)$ hash functions to be evaluated.\footnote{And the same number of bits to access after hashing.}
In dictionary-based filters like cuckoo filters and quotient filters, this is from searching through a near-full table for an open slot to write a fingerprint in.

This brief note describes a Bloom filter in which all operations scale independently of $1/\varepsilon$, with worst-case $O(1)$ operations, first created for Apache Impala in early 2016.\footnote{\url{https://github.com/apache/impala/commit/b35f6d070c8e6b51079f962b448ecc2b0eb74c1a}}\footnote{Rediscovered in 2018 by \cite{ultra-fast}.}
The price paid for this performance is that these filters use a pre-determined number of hash functions, limiting their utility for false positive probabilities outside of $[0.4\%, 19\%]$.
The central ideas are:

\begin{enumerate}
\item Use block Bloom filters to reduce the number of cache lines to access down to one.~\cite{block}
\item Within each block, use a ``split'' Bloom filter that sets one bit in each of several sections, rather than several bits in one section.~\cite{split-bloom}
\item Use eight hash functions in order to fit cleanly into SIMD lanes.\footnote{Four or sixteen would work, too.}
\end{enumerate}

A value is inserted in a split block Bloom filter by first selecting one 256-bit block from the filter by hashing the key once.
Then the key is hashed eight more times to a range of $[0,32)$ using SIMD instructions and multiply-shift universal hashing: $h_{s_i}(x) = \lfloor(s_i \cdot x) / 2^{27}\rfloor$, where $s_i$ are odd seeds.~\cite{multiply-shift}
One bit is set in each of eight contiguous 32-bit lanes within the 256-bit block using the results of the eight hash functions.
Lookup is symmetric; deletions are not supported; code is available at the end of this document.

Each of the central ideas of split block Bloom filters can negatively affect $\varepsilon$ compared to standard Bloom filters.
For instance, in the same space it takes for a split block Bloom filter to support $\varepsilon = 1.0\%$, a standard Bloom filter achieves a false positive rate of $0.63\%$
The false positive rate of split block Bloom filters can be approximated from \cite[Equation 3]{block} and \cite[Section 2.1]{split-bloom}

\[
\sum_{i=0}^\infty P_{256/(m/n)}(i) (1 - (1-8/256)^i)^8
= \sum_{i=0}^\infty P_a(i) (1 - (1-1/32)^i)^8
\]

where $P$ is the Poisson distribution, $n$ is the number of distinct hash values, $m$ is the size of the filter in bits, and $a$ is the average number of distinct hash values per block.
As long as $a \in [20,52]$ ($\varepsilon \in [0.40\%, 19\%]$), the false positive probability of split block Bloom filters is no more than twice that of a standard Bloom filter with the same $n$ and $m$.

In trade-off for this increased $\varepsilon$, we get very high speed.
For instance, comparing against the 8-bit version of cuckoo filters\footnote{This test is replicable from the original cuckoo filter repo, \url{https://github.com/efficient/cuckoofilter}}\footnote{More benchmarks are also shown in \cite{lemire-xor-filter,ribbon-filter,bloom-overtakes}.}:

\begin{tabular}{|r|r|r|r|}
  \hline  & {\bf 100k elements} & {\bf 1M elements} & {\bf 100M elements} \\
  \hline {\bf Size} & 131KB & 1MB & 134M \\
  \hline {\bf Cuckoo insert (M/s)} & 71 & 33 & 14 \\
  \hline {\bf SBBF insert (M/s)} & 416 & 182 & 32 \\
  \hline {\bf Cuckoo lookup (M/s)} & 281 & 139 & 23 \\
  \hline {\bf SBBF lookup (M/s)} & 400 & 186 & 43 \\
  \hline {\bf Cuckoo $\varepsilon$} & 2.37\% & 2.97\% & 2.33\% \\
  \hline {\bf SBBF $\varepsilon$} & 1.03\% & 2.74\% & 0.91\% \\
  \hline
\end{tabular}

Because of their per-slot metadata, cuckoo filters and quotient filters mainly shine at false positive probabilities less than 0.5\%.
For higher probabilities like those in this table, which are in the target range for Impala's use cases, split block Bloom filters are appropriate, even if not the theoretically optimal.
Split block Bloom filters are now also used in StarRocks, Apache Arrow, Apache Kudu, and Apache Parquet.
The filter is available in a standalone package at \url{https://github.com/jbapple/libfilter}.

See the figure below for a cut-down but working C version of split block Bloom filters.

\section*{Acknowledgments}
Thank you to Daniel Lemire for helpful discussions and inspiring questions.

\bibliographystyle{alpha}
\bibliography{doc}

\appendix{}
\begin{figure}
  \begin{framed}
\begin{verbatim}
#include <immintrin.h>
#include <stdint.h>

// Take a hash value and get the block to access within a filter with
// num_buckets buckets.
uint64_t block_index(const uint64_t hash, const uint32_t num_buckets) {
  return ((hash >> 32) * num_buckets) >> 32;
}

// Takes a hash value and creates a mask with one bit set in each 32-bit lane.
// These are the bits to set or check when accessing the block.
__m256i make_mask(uint32_t hash) {
  const __m256i ones = _mm256_set1_epi32(1);
  // Set eight odd constants for multiply-shift hashing
  const __m256i rehash = {INT64_C(0x47b6137b) << 32 | 0x44974d91,
                          INT64_C(0x8824ad5b) << 32 | 0xa2b7289d,
                          INT64_C(0x705495c7) << 32 | 0x2df1424b,
                          INT64_C(0x9efc4947) << 32 | 0x5c6bfb31};
  __m256i hash_data = _mm256_set1_epi32(hash);
  hash_data = _mm256_mullo_epi32(rehash, hash_data);
  // Shift all data right, reducing the hash values from 32 bits to five bits.
  // Those five bits represent an index in [0, 31)
  hash_data = _mm256_srli_epi32(hash_data, 32 - 5);
  // Set a bit in each lane based on using the [0, 32) data as shift values.
  return _mm256_sllv_epi32(ones, hash_data);
}


void add_hash(uint64_t hash, uint32_t num_buckets, __m256i filter[]) {
  const uint64_t bucket_idx = block_index(hash, num_buckets);
  const __m256i mask = make_mask(hash);
  __m256i *bucket = &filter[bucket_idx];
  // or the mask into the existing bucket
  _mm256_store_si256(bucket, _mm256_or_si256(*bucket, mask));
}

_Bool find_hash(uint64_t hash, uint32_t num_buckets, const __m256i filter[]) {
  const uint64_t bucket_idx = block_index(hash, num_buckets);
  const __m256i mask = make_mask(hash);
  const __m256i *bucket = &filter[bucket_idx];
  // checks if all the bits in mask are also set in *bucket. Scalar
  // equivalent: (~bucket & mask) == 0
  return _mm256_testc_si256(*bucket, mask);
}
\end{verbatim}
  \end{framed}
\end{figure}

\end{document}